\documentclass[aps,reprint,superscriptaddress]{revtex4-1}

\usepackage{graphicx}
\usepackage{dcolumn}
\usepackage{bm}
\usepackage{siunitx}
\usepackage{hyperref}
\usepackage{color}
\usepackage[dvipsnames]{xcolor}
\usepackage{float}

\hypersetup{
colorlinks=true,
urlcolor=blue,
citecolor=black,
linkcolor=black
}

\begin{document}

\title
{
Strong mechanically-induced effects in DC current-biased\\ suspended Josephson junctions
}

\author{Thomas McDermott}
\affiliation
{
School of Physics and Astronomy, University of Exeter, EX4 4QL, Exeter, United Kingdom
}
\author{Hai-Yao Deng}
\affiliation
{
School of Physics and Astronomy, University of Exeter, EX4 4QL, Exeter, United Kingdom
}
\author{Andreas Isacsson}
\affiliation
{
Department of Physics, Chalmers University of Technology, SE-412 96, G\"oteborg, Sweden
}
\author{Eros Mariani}
\affiliation
{
School of Physics and Astronomy, University of Exeter, EX4 4QL, Exeter, United Kingdom
}

\date{\today}

\begin{abstract}
Superconductivity is a result of quantum coherence at macroscopic scales. Two superconductors separated by a metallic or insulating weak link exhibit the AC Josephson effect - the conversion of a DC voltage bias into an AC supercurrent. This current may be used to activate mechanical oscillations in a suspended weak link. As the DC voltage bias condition is remarkably difficult to achieve in experiments, here we analyse theoretically how the Josephson effect can be exploited to activate and detect mechanical oscillations in the experimentally relevant condition with \emph{purely DC current bias}. We unveil for the first time how changing the strength of the electromechanical coupling results in two qualitatively different regimes showing dramatic effects of the oscillations on the DC current-voltage characteristic of the device. These include the apperance of Shapiro-like plateaux for weak coupling and a sudden mechanically-induced retrapping for strong coupling. Our predictions, measurable in state of the art experimental setups, allow the determination of the frequency and quality factor of the resonator using DC only techniques.
\end{abstract}
\pacs{}

\maketitle

\section{Introduction}

Superconductivity is a macroscopic quantum phenomenon in which an electrical current flows without dissipation. This supercurrent can also flow between two superconductors separated by a metallic or insulating weak link, a phenomenon known as the Josephson effect and attributed to quantum tunneling of electron pairs \cite{josephson}. Josephson also predicted that, if a DC voltage bias is maintained across such a (Josephson) junction, the supercurrent then alternates due to interference between the macroscopic wave functions of the two superconductors. Soon after Josephson's work it was realized that, if the current in the junction is coupled to external AC radiation \cite{shapiro} or internal electromagnetic resonances in the weak link \cite{coonfiske,baberschke}, new constant-voltage steps in the current-voltage (I-V) characteristic emerge when the AC supercurrent frequency matches a multiple of the intrinsic resonator frequency. Recently, this scenario has been extended, both theoretically and experimentally, to the case in which the weak link itself acts as a mechanical resonator  \cite{zhou,buks,blencowe,buks2,zhu,sonne1,sonne2,sonne3,padurariu,marchenkov,keijzers,etaki,etakiSSO,kretinin}. It has been theoretically predicted that the AC supercurrent can pump mechanical oscillations in the resonator due to the coupling between electronic and mechanical degrees of freedom \cite{zhu,sonne1,sonne2,sonne3,padurariu}. Experimental signatures of the excitation of mechanical resonances in vibrating weak-links have been reported in atomic scale oscillators produced in break junctions \cite{marchenkov}, in torsional SQUID resonators \cite{etaki} and in suspended nanowires \cite{kretinin}.
These early observations testify the potential of current experimental setups to fully explore novel electromechanical effects in the context of superconductivity.

The interplay between electronic currents and vibrations has been explored extensively in the context of quantum transport through non-superconducting nanoelectromechanical systems (NEMS). The most striking manifestations of the excitation of mechanical vibrations by the electronic currents through the NEMS are the appearance of vibrational sidebands in the I-V characteristic, accompanied by a dramatic suppression of current at low bias (Franck-Condon blockade) when the electromechanical coupling is strong enough \cite{koch,koch2,sapmaz,leturcq}. The analogous effects associated with the strong coupling regime in suspended Josephson junctions have never been explored so far.
Moreover, the theoretical analysis of the I-V characteristic of suspended Josephson junctions has so far been limited to the voltage bias case \cite{zhu,sonne1,sonne2,padurariu}. While this is a convenient theoretical approach, it has serious limitations in addressing the response of experimental devices due to a major constraint in the operation of Josephson junctions: their small impedance in comparison to that of the external circuit makes them invariably operate in the current bias regime even if one attempts to maintain a fixed voltage bias across them \cite{likharev}. 

In this theoretical paper we address these open issues for the first time. By means of numerical as well as analytical investigations, we show how existing experimental setups can be used to induce and detect high frequency mechanical oscillations in suspended weak links using \emph{purely DC current bias} conditions. Exploiting a setup that allows one to tune the coupling strength between electronic and mechanical degrees of freedom, we explore the Josephson effect in regimes that have not been studied before, revealing several qualitatively new features in the DC I-V characteristic. Among them we analyze the Shapiro steps-like features that appear for weak electromechanical coupling. In contrast, we unveil that for strong coupling such steps collapse to a zero voltage state with a sudden mechanical-induced retrapping due to energy being subtracted from the electronic system by the oscillations. Accessing the strong coupling regime results in a dramatic shift in the retrapping current of order 50\%. We reveal how the crossover between the weak and strong coupling regimes is intimately related to the quality factor of the resonator. Remarkably, our predictions suffice to unveil all the fundamental properties of the resonator, i.e. not just its proper frequency but also the quality factor, allowing their direct experimental measurements by simply recording the DC I-V curve without any additional measurement setup. The experimental realisation of our theoretical proposal in state of the art devices is carefully discussed.  

\section{Model}

\begin{figure}
    \includegraphics[width=75mm]{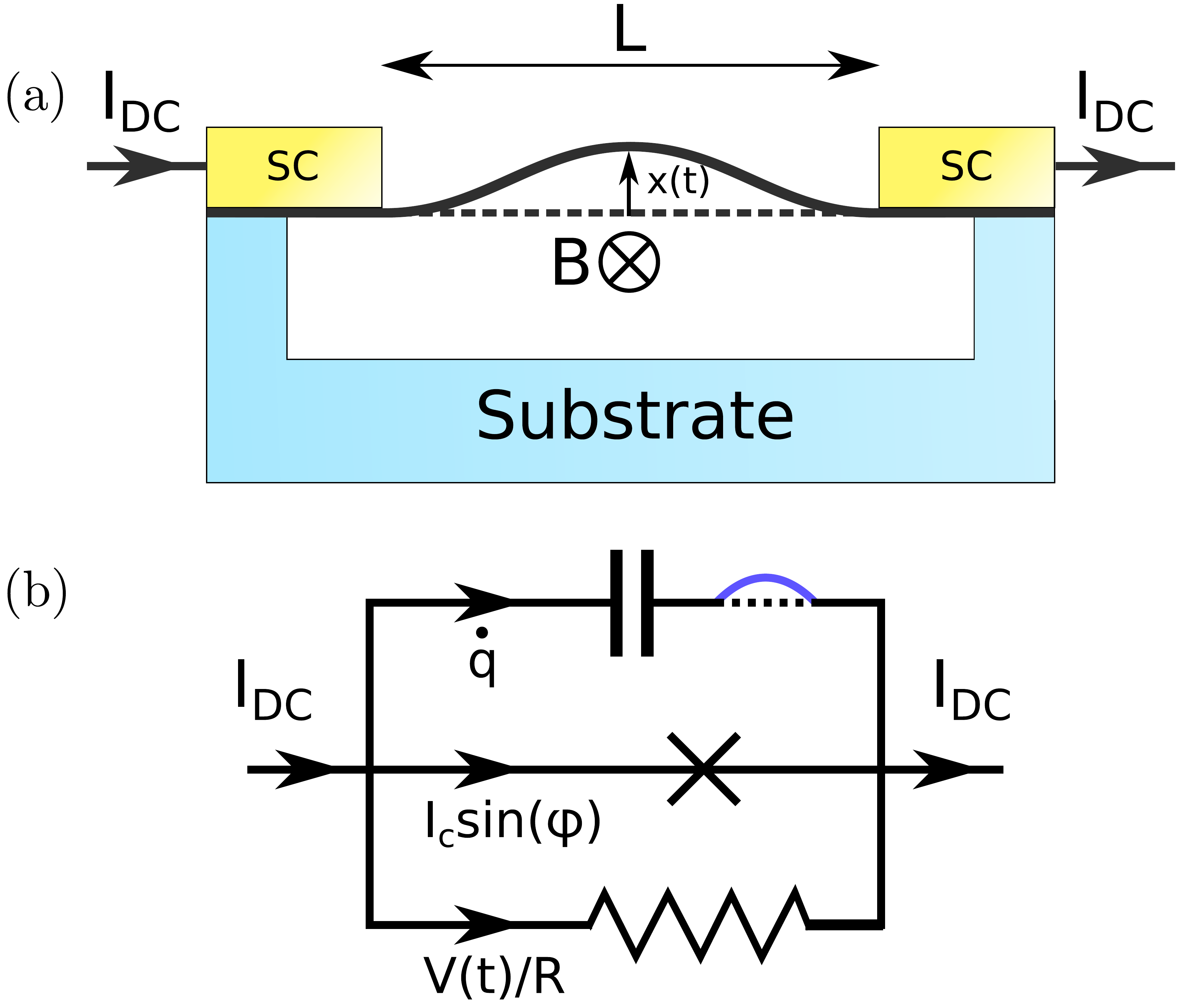}
    \caption{(a) Electromechanical resonator suspended between two superconducting contacts above a substrate. The system is biased with a current $I_\text{DC}^{}$ and vibrates in the presence of an in-plane magnetic field. A constant back-gate voltage applied to the substrate can be used to tune the mechanical resonant frequency. (b) Equivalent circuit of the RCSJ model with mechanical oscillations. Current may flow through either resistive, capacitive or supercurrent channels. Mechanical oscillations produce an extra voltage across the effective capacitor.}
    \label{device}
\end{figure}

We analyse theoretically a nanomechanical resonator of length $L$ suspended between two superconducting contacts and biased by a DC current $I_\text{DC}^{}$, see Fig.\,\ref{device}(a). The suspended resonator is subject to an in-plane magnetic field $B$ that allows one to tune the coupling between the electronic current and the fundamental flexural deformation mode via the Lorentz force. This rather standard setup can be realised experimentally in various ways using different suspended resonators, e.g. nanowires \cite{kretinin}, one dimensional carbon nanotubes (CNTs) \cite{peng,garcia,huttel,laird,moser} or ultra-thin two-dimensional materials like graphene \cite{bunch,chen} and transition metal dichalcogenides \cite{tmd}. The proposed device behaves as a Josephson weak link with a supercurrent $I_\text{c}^{} \sin \varphi$ flowing between the contacts, where $I_\text{c}^{}$ is the critical current of the weak link and $\varphi$ is the gauge invariant phase difference between the macroscopic wave functions in the two contacts. The Josephson relation demands that the frequency $d\varphi/dt$ is related to the voltage $V$ across the weak link by 
\begin{equation}
    \frac{d\varphi}{dt} =\frac{2eV}{\hbar}.
\label{josephson}
\end{equation} 
\noindent In the current bias regime that we consider here, this voltage is a dynamic variable whose value is determined by the combination of the working parameters of the Josephson weak link as well as its mechanical motion. In the next sections we show how, despite a purely DC current bias, an essentially constant voltage $V$ can emerge. This can be used to match the supercurrent frequency in Eq. (\ref{josephson}) to the resonant frequency of the resonator $\omega_0^{}$, leading to forced amplification of the oscillations. 

\subsection{Uncoupled case: $B=0$}

We describe the current flow by means of the standard resistively and capacitively shunted junction (RCSJ) model \cite{tinkham}, by which the current is split into resistive, capacitive and supercurrent channels, see Fig.\,\ref{device}(b). When the Josephson dynamics and the mechanical oscillations are uncoupled (i.e. for $B=0$) current conservation reads
\begin{equation}
    I_\text{DC}^{} = I_\text{c} \sin \varphi + \frac{V}{R} + C \frac{dV}{dt},
\label{rcsjuncoupled0}
\end{equation}
\noindent where $R$ and $C$ are the effective resistance and capacitance of the weak link. Upon using the Josephson relation (\ref{josephson}), Eq.\,(\ref{rcsjuncoupled0}) translates into a differential equation for the phase difference $\varphi$
\begin{equation}
    \frac{I_\text{DC}^{}}{I_\text{c}^{}} = \sin \varphi + \frac{1}{\omega_\text{c}^{}}\frac{d\varphi}{dt} + \frac{\beta_\text{c}^{}}{\omega_\text{c}^2}\frac{d^2\varphi}{dt^2}.
\label{rcsjuncoupled}
\end{equation}
\noindent Here $\omega_\text{c}^{}=2eI_\text{c}^{} R/\hbar$ is the characteristic frequency of the supercurrent corresponding to a voltage bias $I^{}_{\rm c}R$ and $\beta_\text{c}^{} = \omega_\text{c}^{} R C$ is the Stewart-McCumber parameter that can be expressed as the ratio $\beta_\text{c}^{} = |Z_\text{R}^{}|/|Z_\text{C}^{}|$, where $Z_R^{} = R$ and $Z_C^{} = -i/\omega_c^{} C$ are the impedances of the resistive and capacitive channels at the frequency $\omega_\text{c}$, respectively. In Eq.\,(\ref{rcsjuncoupled}) the first and second terms on the RHS describe the supercurrent and resistive current, both of which involve the flow of electrons through the weak link. In contrast, the third term describes the displacement current $dq/dt$ (with $q(t)$ the charge on the effective capacitor) due to charging effects on the capacitor and not associated with any electronic current. 

In the finite voltage state the supercurrent $I_\text{c} \sin \varphi$ is oscillatory due to the Josephson relation (\ref{josephson}), and must be compensated by either the resistive or capacitive channels in order for the total current, $I_\text{DC}^{}$, to be constant. In the overdamped regime, $\beta_\text{c} \ll 1$, the impedance of the capacitive channel dominates the resistive one leading to a negligible displacement current through the capacitor. The AC supercurrent is thus compensated by a resistive current (and therefore voltage) which becomes highly oscillatory. In contrast, in the underdamped regime, $\beta_\text{c} \gg 1$, the impedance of the resistive channel dominates the capacitive one so that the supercurrent is compensated by the displacement current leaving an almost constant voltage with small fluctuations. In this way an essentially constant voltage can be achieved even using a DC current bias setup as long as the weak link is underdamped. As the total \emph{electronic} current passing through the weak link is $I_\text{DC}^{}-dq/dt$, in the overdamped regime this current is DC whereas in the underdamped case it is strongly AC. We will exploit this AC current coupled with the in plane magnetic field to activate mechanical resonances in underdamped Josephson weak links. We note that typical graphene and CNT based suspended Josephson weak links operate in this regime with $\beta_\text{c} \sim 10-100$ \cite{heersche,mizuno,herrero}.

\begin{figure}
    \includegraphics[width=75mm]{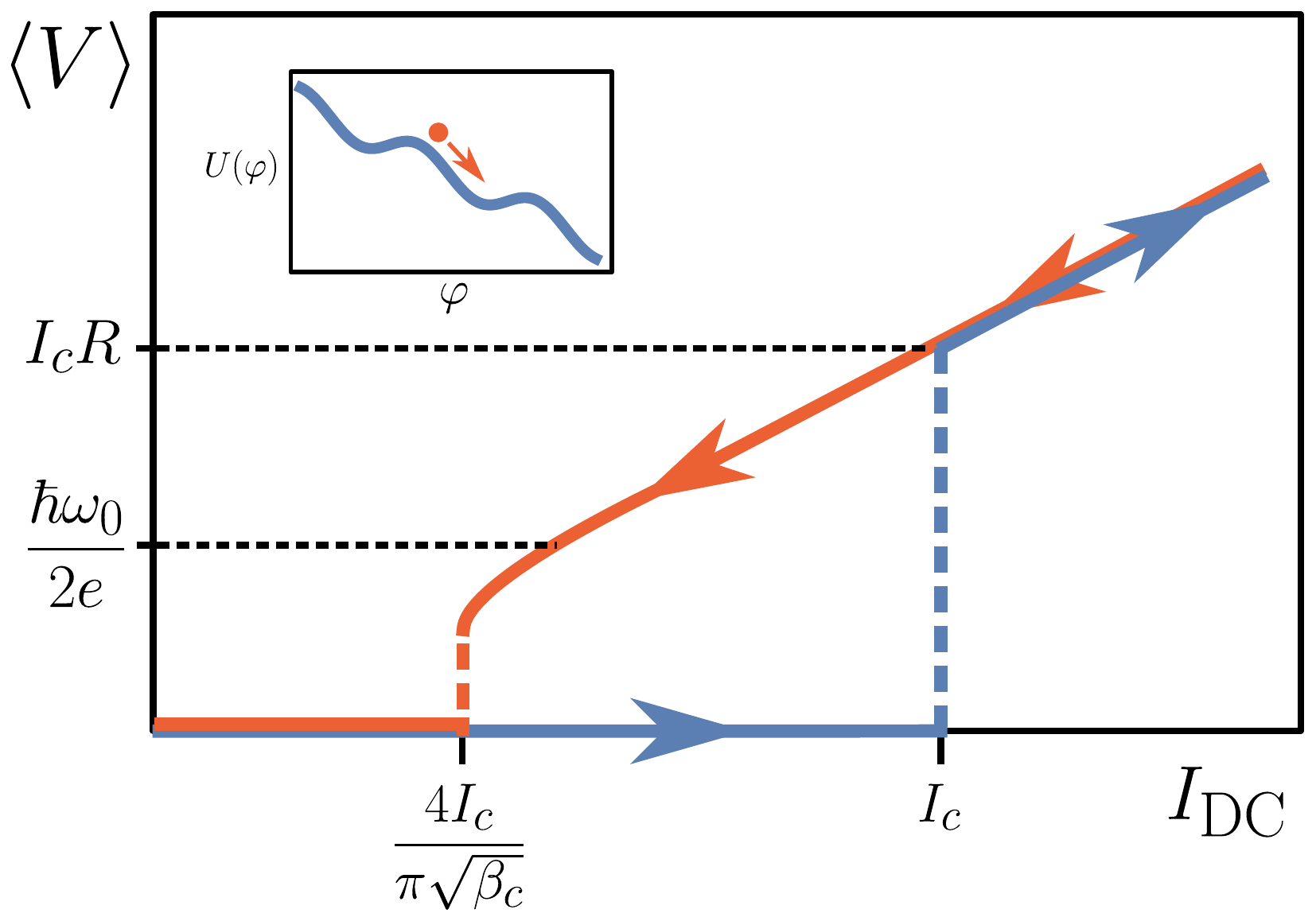}
    \caption{Sketch of the DC voltage as a function of the applied current $I_\text{DC}^{}$ for the uncoupled case, Eq. (\ref{rcsjuncoupled}), for an underdamped ($\beta_\text{c}^{} \gg 1$) weak link. Hysteresis is present so that different voltages are measured on the increasing (blue line) and decreasing (orange line) current paths. On the increasing path the system is trapped in a zero voltage state until $I_\text{DC}^{}$ exceeds $I_\text{c}^{}$ where there is an abrupt transition into an Ohmic running state $\langle V \rangle \approx I R$. Upon decreasing the current again, the system remains in the running state until one reaches a retrapping current $4I_\text{c}^{}/\pi \sqrt{\beta_\text{c}^{}}$. The voltage associated with the mechanical resonance $\hbar \omega_0^{} / 2e$ is typically much less than $I_\text{c}^{} R$ so that it can only be reached on the decreasing current path. A sketch of the effective `tilted washboard' potential of the mechanical analog to Eq. (\ref{rcsjuncoupled}) is shown in the inset.}
    \label{IVuncoupled}
\end{figure}

The Josephson dynamics can be interpreted physically in terms of a mechanical analogy that will be used often in the remainder of the paper. In fact Eq. (\ref{rcsjuncoupled}) describes the motion of a particle of mass $(\hbar/2e)^2 C$ moving along the $\varphi$ axis in the `tilted washboard' potential
\begin{equation}
    U(\varphi) = -\frac{\hbar I_\text{c}}{2e} \left(\cos \varphi + \frac{I_\text{DC}^{}}{I_\text{c}} \varphi\right),
\label{washboard}
\end{equation}
\noindent under the effect of a drag force $(\hbar/2e)^2(1/R)d\varphi/dt$, as shown in the inset of Fig.\,\ref{IVuncoupled}. In this analogy the impressed current $I_\text{DC}^{}$ is proportional to the tilt of the washboard potential, while $\beta_\text{c}$ is inversely proportional to the damping of the system. In addition, the voltage $V$ and capacitance $C$ relate to the velocity and mass of the particle, respectively. By measuring the DC voltage $\langle V(t) \rangle$ (where $\langle...\rangle$ denotes time averaging) across the weak link as a function of the impressed current $I_\text{DC}^{}$, one obtains the I-V characteristic as a key experimental signature of the dynamics of the system. In the underdamped regime the I-V curve is highly hysteretic (see sketch in Fig.\,\ref{IVuncoupled}) and we must distinguish between the curves obtained on the increasing and decreasing current paths. As the impressed current $I_\text{DC}^{}$ is increased above the critical current $I_\text{c}$ there is an abrupt transition from the `trapped state' $\langle V \rangle = 0$ where the washboard particle is trapped in a potential well, to a `running state' $\langle V \rangle \approx I R$ where the local potential minima vanish and the particle rolls down the washboard reaching its terminal velocity. If one then decreases $I_\text{DC}^{}$ below $I_\text{c}$ again, the system does not become immediately retrapped but remains in the running state until the current is decreased below a retrapping current $I_\text{r} = 4 I_\text{c} / \pi \sqrt{\beta_\text{c}}$ \cite{tinkham} due to the inertia of the particle, or, in terms of the electronic system, due to the charging of the capacitor.

To achieve mechanical resonance in the weak link, we require that the supercurrent frequency $d\varphi/dt$ matches the resonant frequency $\omega_0$, i.e. $\langle V \rangle = V_0 = \hbar\omega_0/2e$. As $\omega_0$ is typically a few orders of magnitude smaller than $\omega_\text{c}$ (both frequencies are sample specific but typical estimates yield  $\omega_0/\omega_\text{c} \approx 1 \times 10^{-3}$ for graphene devices \cite{heersche,mizuno,bunch,chen} and $\omega_0/\omega_\text{c} \approx 0.1$ for CNTs \cite{cleuziou,herrero,peng,garcia,huttel,laird,moser}), this resonance condition is difficult to achieve on the increasing current path. However, if $\beta_\text{c}$ is large enough, thanks to the hysteresis of the weak link, we may instead choose the decreasing current path, reaching a voltage close enough to the resonance, see Fig.\,\ref{IVuncoupled}. This is only possible if $I_\text{r}/I_\text{c}$ is less than $\omega_0/\omega_c$, i.e.
\begin{equation}
    \beta_\text{c}^{} > \left( \frac{4}{\pi}\frac{\omega_\text{c}}{\omega_0}\right)^{2}_{}\; .
\label{condition}
\end{equation}
\noindent This condition must be satisfied for the proposed activation mechanism to work. For CNT weak links this corresponds to $\beta_\text{c} \gtrsim 100$, which is experimentally achievable with state of the art devices.

\subsection{Coupled case: $B \neq 0$}

The presence of an in-plane magnetic field has two major effects. Firstly, it generates a Lorentz force on the electronic currents which induces oscillations in the NEMS. These in turn produce an additional voltage contribution across the weak link that redistributes the current through the channels. The total voltage is given by
\begin{equation}
    \frac{\hbar}{2e}\frac{d\varphi}{dt} = V = \frac{q}{C}-B L \frac{dx}{dt},    
\end{equation}
\noindent where $x(t)$ is the oscillator displacement. This equation is a direct consequence of gauge invarance. It shows that mechanical oscillations alter the voltage across the capacitor and the capacitive current $dq/dt$. The equation of current conservation now becomes
\begin{equation}
    I_\text{DC}^{} = I_\text{c} \sin \varphi + \frac{V}{R} + C \left( \frac{dV}{dt} + BL \frac{d^2x}{dt^2} \right).
\label{rcsjdimensions}
\end{equation}
\noindent In comparison with the uncoupled case, an extra current emerges resulting directly from the electromechanical coupling. It is through this extra term that the oscillator affects the I-V characteristic of the weak link and allows the oscillations to be detected. In particular, during resonance this extra current oscillates with a frequency $\omega_0$. As this scenario is analogous to the case of a Josephson weak link biased by an AC current \cite{tinkham}, we may expect a Shapiro plateau to develop at a voltage $V_0$ when resonances are induced. Additional resonances can also be induced at voltages $n V_0$ where $n$ is an integer, leading to higher order Shapiro steps.

The flexural mode of the suspended weak link is modelled mechanically as a simple harmonic oscillator with mass $M$, proper frequency $\omega_0$, damping coefficient $\Gamma$ and quality factor $Q=\omega_0 / \Gamma$. As to be seen later, anharmonic effects are irrelevant in this work as only small amplitude mechanical oscillations will be activated. Taking into account the Lorentz force exerted on the electronic currents by the magnetic field, the equation of motion of the oscillator can be written as
\begin{equation}
    \frac{d^2x}{dt^2} +2\Gamma \frac{dx}{dt} + \omega_0^2 x = \frac{BL}{M} \left(I_\text{c} \sin \varphi + \frac{V}{R}\right).
\label{mechdimensions}
\end{equation}
\noindent Substituting Eq. (\ref{rcsjdimensions}) into Eq. (\ref{mechdimensions}), we obtain
\begin{equation}
    \left(1+\frac{B^2_{}}{B_{0}^2}\right) \frac{d^2x}{dt^2} + 2\Gamma \frac{dx}{dt} + \omega_0^2 x = \frac{BL}{M} \left( I_\text{DC}^{} - C \frac{dV}{dt} \right).
\label{mechdimensions2}
\end{equation}
\noindent The electromechanical coupling produces a correction to the effective oscillator mass of the form $M(B/B_{0}^{})^2$, where we introduced the magnetic field scale $B_0^{}=(1/L)\sqrt{M/C}$. The strength of the electromechanical coupling can then be expressed in terms of the dimensionless parameter $\mu = B/B_0$. Similarly, we introduce dimensionless quantities for the current $i_\text{DC}^{} = I_\text{DC}^{}/I_\text{c}$, time $\tau = \omega_0 t$ and oscillator displacement $a = x/x_0 - i_\text{DC}^{}$. Here $ x_0 = B_0 I_\text{c} L / M \omega_0^2$ is the displacement at which the restoring force equals the magnetic force scale $B_0 I_\text{c} L$ and we have subtracted the small constant displacement arising from the force $BI_\text{DC}^{}L$.

In terms of these dimensionless quantities, Eqs. (\ref{rcsjdimensions}) and (\ref{mechdimensions2}) can be rewritten as
\begin{eqnarray}
    && i_{\text{DC}} = \sin \varphi + \beta_1 \dot{\varphi} + \beta_2 \ddot{\varphi} + \mu \ddot{a},
\label{rcsj}\\
    && (1+\mu^2)\ddot{a} + \frac{2}{Q} \dot{a} + a = - \mu \beta_2 \ddot{\varphi}.
\label{mech}
\end{eqnarray}
\noindent Here $\beta_1 = \omega_0/\omega_\text{c}$, $\beta_2 = \beta_1^2 \beta_\text{c}$ and $\dot{f}$ refers to the derivative of $f$ with respect to $\tau$. Equations (\ref{rcsj}) and (\ref{mech}) capture the essential aspects of the coupled Josephson-mechanical system. 

Due to the highly non-linear nature of Eqs. (\ref{rcsj}) and (\ref{mech}), their solutions are expected to display strong dependence on initial conditions. A general exact analytical solution cannot be achieved. In what follows, we first numerically solve the equations to establish the characteristic I-V curve. The properties of the numerical solutions will then suggest an ansatz for a semi-analytical treatment of the problem leading to a good agreement with the numerically established I-V curve.

\section{Numerical Solution}

\begin{figure*}
	\includegraphics[width=160mm]{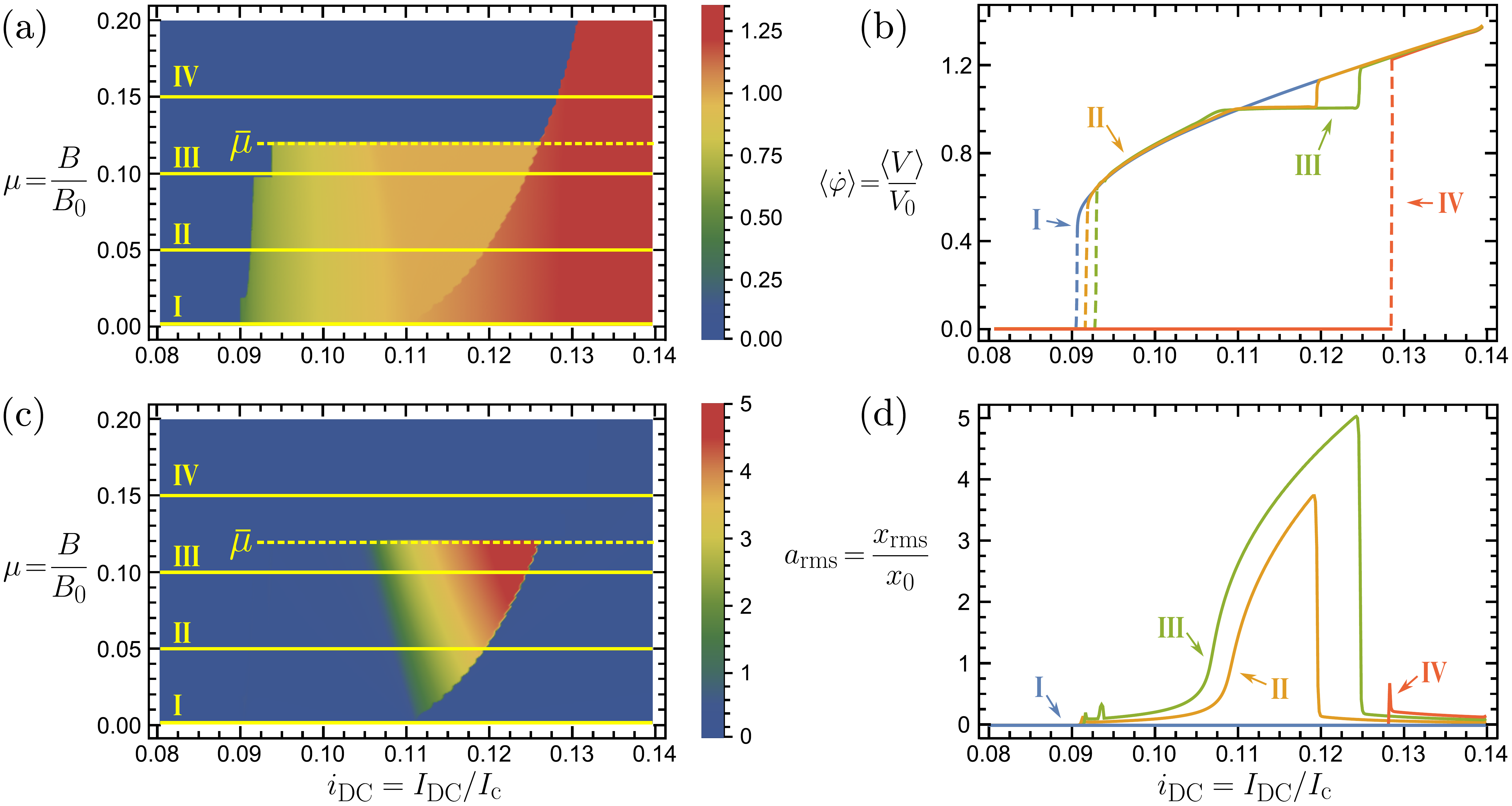}
	\caption{ DC voltage and mechanical oscillation amplitude as functions of the applied current $I_\text{DC}^{}$ and magnetic field $B$, obtained by numerically solving Eqs. (\ref{rcsj}) and (\ref{mech}) with $\beta_1 = 0.1$, $\beta_2 = 2$, $Q=10^3$. Typical scales for relevant experimental parameters (see methods) are $I_\text{c} \approx 10 \, \si{\nano \ampere}$, $\omega_0 \approx 1 \si{\giga \hertz}$, $V_0 \approx 0.3 \, \si{\micro \volt}$, $B_0 \approx 10 \si{\tesla}$, $x_0 \approx 10 \, \si{\pico \metre}$. Key features exhibited in this figure are present for any choice of parameters as long as Eq.\, (\ref{condition}) is satisfied. (a) Colour plot showing the I-V curve along decreasing current paths as a function of the coupling strength $\mu = B/B_0$. (b) I-V curves for specific values of coupling ($\mu = 0, 0.05, 0.10, 0.15$) corresponding to the cuts I-IV in panel (a) respectively, showing clear progression from the uncoupled case ($\mu = 0$), to a plateau appearing at the resonant frequency ($\mu = 0.05, 0.10$), to the sudden retrapping when $\mu > \bar{\mu}$ ($\mu = 0.15$). (c) Colour plot of the root mean square (RMS) dimensionless displacement $a_\text{rms} = \sqrt{\langle a^2 \rangle}$. Comparison with panel (a) shows that the voltage plateau coincides with the mechanical resonance. (d) Oscillation amplitude versus current for specific values of coupling ($\mu = 0, 0.05, 0.10, 0.15$) corresponding to the cuts I-IV in panel (c) respectively. The amplitude $x$ of the resonant oscillations is of the order of a few tens of $\si{pm}$ which, for a typical resonator with $L\sim 1\,\si{\micro \metre}$ corresponds to a strain of order $(x/L)^{2}_{}\sim 10^{-10}_{}$, thus justifying neglecting non-linearities in our model.}
	\label{numerical}
\end{figure*}

\begin{figure}
	\includegraphics[width=75mm]{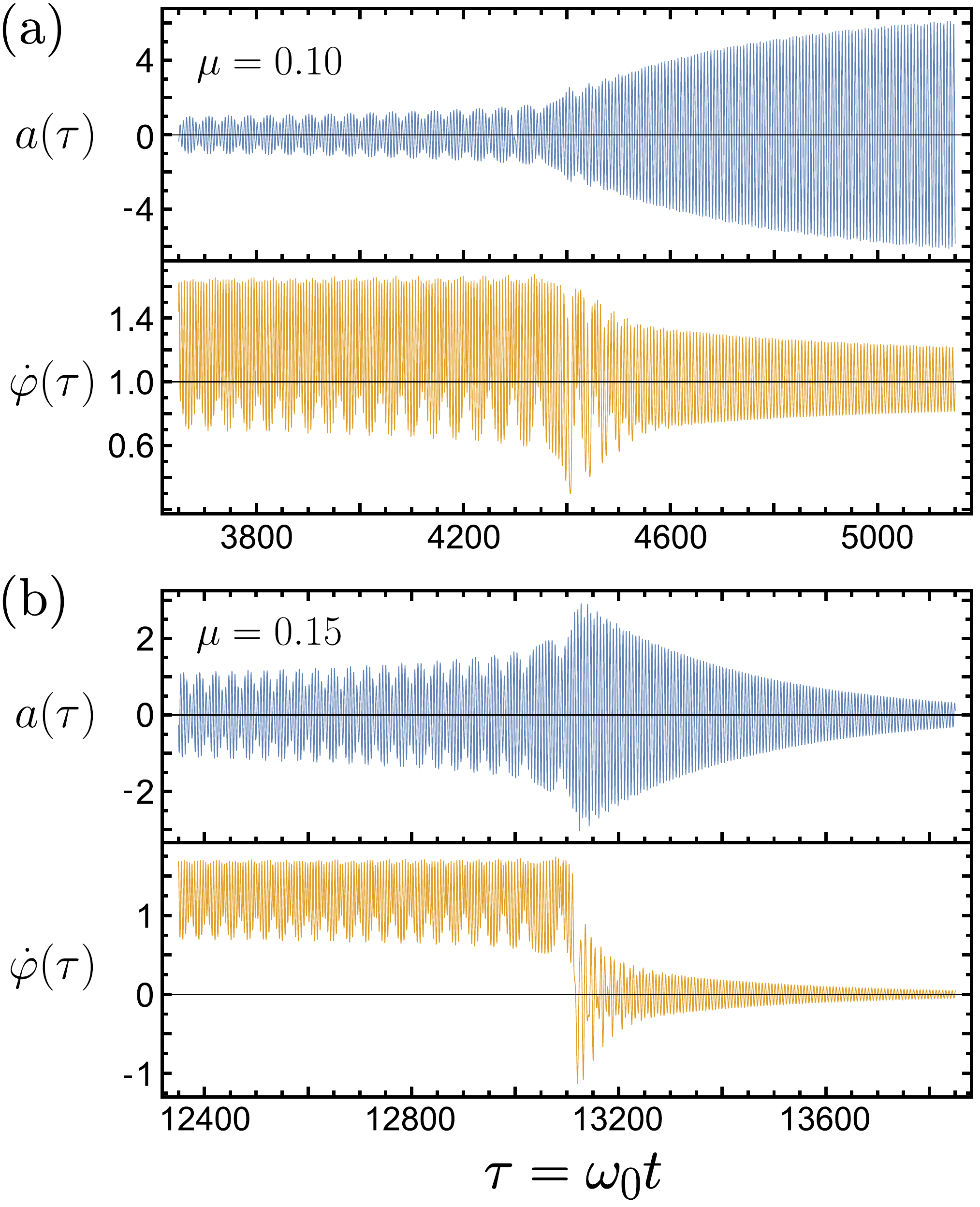}
	\caption{Numerical solutions of Eqs. (\ref{rcsj}) and (\ref{mech}) with $\beta_1 = 0.1$, $\beta_2 = 2$ and $Q=10^3$ for (a) $\mu=0.10$ and (b) $\mu=0.15$. At $\tau=0$ in both cases, the current has been lowered to reach the plateau region. After a long transient period the system enters a resonant state and mechanical oscillations increase in amplitude. The subsequent evolution depends on whether the coupling $\mu$ is above or below the critical value $\bar{\mu} \approx 0.12$. For $\mu = 0.10$, the oscillations continue to increase but eventually saturate, while the supercurrent frequency is locked at the resonant frequency, $\langle \dot{\varphi}\rangle = 1$. However, for $\mu = 0.15$, the oscillations increase too rapidly and cannot be maintained; instead the system is abruptly retrapped into the zero voltage state. The supercurrent frequency rapidly goes to zero and oscillations decay.}
	\label{timeevolution}
\end{figure}

We numerically solve Eqs. (\ref{rcsj}) and (\ref{mech}) using a fourth order Runge-Kutta method. The experimentally tunable parameters are the input current $i_\text{DC}^{}$ and the coupling parameter $\mu$, while the directly measurable quantity is the average voltage, $\langle \dot{\varphi} \rangle = V/V_0$. All parameters other than $i_\text{DC}^{}$ and $\mu$ are sample specific. The results shown in Fig.\,\ref{numerical} were obtained with $\beta_1 = 0.1$, $\beta_2 = \beta_1^2 \beta_c = 2$ which are of the same order as those found in real graphene and CNT mechanical resonators \cite{bunch,chen,peng,garcia,huttel,laird,moser} and in Josephson weak links \cite{heersche,mizuno,herrero,cleuziou}, while the quality factor is chosen to be $Q=10^3$. In experimental samples the quality factors fluctuate a lot, reaching values of $Q$ up to $10^6$ \cite{moser}. In the next section, we show that $\mu \propto Q^{-1/3}$ and thus a change in $Q$ simply rescales the magnetic field at which the coupling effects may be observed. We have verified that the general features of our analysis are present for a wide range of parameters as long as the resonance region can be reached, i.e. if Eq. (\ref{condition}) is satisfied, which in terms of $\beta_2$ simply reads $\beta_2 > (4/\pi)^2 \approx 1.62$. The dimensional scales introduced in the previous section may be estimated from experimental parameters (see methods), yielding $I_\text{c} \approx 10 \, \si{\nano\ampere}$, $\omega_0 \approx 1 \, \si{\giga \hertz}$, $V_0 \approx 0.3 \, \si{\micro\volt}$, $B_0\approx 10 \, \si{\tesla}$ and $x_0 \approx 10 \, \si{\pico\meter}$. While $B_0$ is rather large, we show below that the actual magnetic fields required to observe strong electromechanical effects are much lower.

We employ continuous initial conditions whereby we start with $i_\text{DC}^{}>1$, where the only solution is the running state $\langle \dot{\varphi} \rangle \approx i_\text{DC}^{}/\beta_1$, and gradually decrease $i_\text{DC}$ in small increments, at each point using as initial conditions the results for $a, \varphi, \dot{a}$ and $\dot{\varphi}$ from the simulation for the previous value of $i_\text{DC}^{}$ after transients have decayed. This adequately models the quasi-static process in which the characteristic time scales of the system are much shorter than the time over which the impressed current $i_\text{DC}^{}$ is experimentally varied. This process is then repeated to obtain the I-V curves for different values of $\mu$. We point out that an alternative experiment whereby the magnetic field $\mu$ is varied at fixed $i_\text{DC}$ will access different solutions and thus yield different results from those discussed here.

Fig.\,\ref{numerical} shows the DC voltage and mechanical oscillation amplitude as functions of the applied current $i_\text{DC}^{}$ and magnetic field coupling $\mu$. Panels (a) and (b) show how the I-V curve on the decreasing current path changes with $\mu$. At $\mu=0$ we recover the usual I-V curve in the absence of coupling with a retrapping current $I_\text{r} = 4 I_\text{c}/\pi \sqrt{\beta_\text{c}}$  (see Fig.\,\ref{IVuncoupled}). Upon increasing $\mu$, the retrapping current increases slightly while a Shapiro-like plateau develops at the frequency of the oscillator $\langle \dot{\varphi} \rangle = V/V_0 \approx 1$. The width of this plateau increases with the applied field. Our numerical analysis reveals similar structures around successive integer values of $\langle\dot{\varphi}\rangle$, though with progressively smaller plateau width. The shoulder-like features observed experimentally in the I-V curve in ref.\,\cite{kretinin} can be interpreted as a signature of this effect for higher-frequency in-plane vibrational modes. Fig.\,\ref{numerical}(c) and \ref{numerical}(d) show the root mean square (rms) mechanical displacement $a_\text{rms}=\sqrt{\langle a^2 \rangle}$. The comparison with panels (a) and (b) shows that the resonance of the oscillator coincides with the plateau indicating that the latter is a result of the effective AC current $\mu \ddot{a}$ in Eq. (\ref{rcsj}) due to the coupling to mechanical oscillations. This is highlighted in Fig.\,\ref{timeevolution}(a) where we show the time evolution of the voltage and the mechanical oscillation amplitude as the resonant state is entered. After a long transient the voltage becomes locked to $V_0$, signalling the matching between the supercurrent and resonant frequency leading to a large amplification of the oscillations.

An even more dramatic effect occurs when the coupling $\mu$ exceeds a critical value $\bar{\mu}$ (see Fig.\,\ref{numerical}). In this case, as the current is decreased, instead of the voltage becoming pinned to the plateau, the system suddenly retraps itself into the zero voltage state, giving rise to a dramatic increase of the retrapping current by as much as 50\%. This suppression of voltage in the strong coupling regime has not been explored previously and resembles a Josephson junction analogue of the Franck-Condon blockade observed in quantum transport through NEMS \cite{koch,koch2,sapmaz,leturcq}. Fig.\,\ref{timeevolution}(b) shows the time evolution of the solutions for $\mu > \bar{\mu}$ which reveals more information about the retrapping. When the resonant condition is met the system attempts to enter the resonant state with the amplitude of $a(\tau)$ increasing rapidly. However, as discussed in the analytical treatment below, this sudden mechanical amplification takes away too much energy from the electronic subsystem. The running state thus can no longer be sustained and the system retraps.

Concerning the experimental observation, the predicted effects occur at a typical current and voltage $\beta_1 I_\text{c} \approx 1 \, \si{\nano \ampere}$ and $V_0 \approx 0.3 \, \si{\micro\volt}$ (corresponding to $\omega_0 \approx 1 \si{\giga \hertz}$), respectively. The critical magnetic field inducing the strong coupling regime and the sudden retrapping is $\bar{\mu} B_0$ which is a small fraction of the rather large $B_0$. As shown later $\bar{\mu}$ scales as $Q^{-1/3}$ so that the predicted effect can be observed with magnetic fields as small as $\bar{\mu} B_0 \approx 100 \, \si{m\tesla}$ for $Q=10^6$.

Our numerical analysis also reveals the appearance of two small plateaux-like features in Fig.\,\ref{numerical}(a) at $\mu\simeq 0.02$ and $\mu\simeq 0.1$. These are accompanied by a small amplification of the oscillator amplitude visible in the curve III in Fig.\,\ref{numerical}(d) at around $i_{{\rm DC}}^{}\simeq 0.093$. The origin of these features is still unclear and will be the subject of further investigation.

\section{Analytical Solution}

To peer further into the nature of the I-V curve, in particular the formation of the plateau and the abrupt retrapping for $\mu>\bar{\mu}$, here we develop an analytical approach to Eqs. (\ref{rcsj}) and (\ref{mech}). These may be derived from the Euler-Lagrange equations
\begin{equation}
    \frac{d}{d\tau}\frac{\partial L}{\partial \dot{q_i}}-\frac{\partial L}{\partial q_i}= - \frac{\partial P}{\partial \dot{q_i}},
\end{equation}
\noindent with generalised coordinates $q_i = \{ \varphi, a\}$, where the gauge-invariant Lagrangian $L$ and dissipation function $P$ are given by
\begin{eqnarray*}
    && L = \frac{1}{2}\beta_2 \left(\dot{\varphi}+\frac{\mu}{\beta_2} \dot{a} \right)^2 + \cos \varphi + i_\text{DC}^{} \varphi + \frac{1}{2 \beta_2} \left(\dot{a}^2-a^2\right)\label{lagrangian},\\
    && P = \frac{1}{2} \beta_1 \dot{\varphi}^2 + \frac{1}{Q \beta_2}\dot{a}^2.
\label{dissipation}
\end{eqnarray*}
The Hamiltonian is $H = E - i_\text{DC}^{} \varphi$ where the total energy of the system $E$ is given by
\begin{equation}
    E = \underbrace{\frac{1}{2}\beta_2 \dot{\varphi}^2 - \cos \varphi}_{E_\varphi} + \underbrace{\frac{1}{2\beta_2}(\dot{a}^2(1+\mu^2)+a^2)}_{E_m} + \underbrace{\mu \dot{a}\dot{\varphi}}_{E_c}.
\label{energyeq}
\end{equation}
\noindent Here $E$ is measured in units of $\hbar I_\text{c} / 2 e$ and it can be split into $E_\varphi$, $E_m$ and $E_c$ corresponding to the energy of the electronic and mechanical subsystems and the coupling energy between them,  respectively. The rate of energy change is given by
\begin{equation}
    \frac{dE}{d\tau} = i_\text{DC}^{} \dot{\varphi} - \beta_1 \dot{\varphi}^2- \frac{2}{Q \beta_2}\dot{a}^2,
\label{energyeq2}
\end{equation}
\noindent where the first term on the right hand side is the power supplied by the external current while the second and third terms describe energy losses due to Joule heating and the intrinsic damping of the resonator. When the system reaches a steady state the supplied energy must be completely dissipated on average, i.e. 
\begin{equation}
    i_\text{DC} \langle \dot{\varphi} \rangle = \beta_1 \langle \dot{\varphi}^2 \rangle + \frac{2}{Q\beta_2} \langle \dot{a}^2 \rangle.
\label{energyeq3}
\end{equation}
\noindent This equation has been thoroughly checked numerically. In the uncoupled case, it simply leads to the Ohmic solution $\langle \dot{\varphi} \rangle \approx i_\text{DC}^{}/\beta_1$. Upon coupling to mechanical oscillations, the intrinsic damping of the oscillator must also be considered. Energy is transferred from the electronic subsystem to mechanical oscillations leading to measurable effects on the total voltage across the device.

The numerical solutions reveal that $\varphi(\tau)$ takes on the following simple form whenever the system reaches a steady state,
\begin{equation}
    \varphi = \varphi_0 + \omega \tau - \frac{g}{\omega} \cos(\omega \tau),
\label{ansatz1}
\end{equation}
\noindent where $\varphi_0$ is a constant phase that is not related to initial conditions but acquired during evolution, while $g$ and $\omega$ are parameters to be determined self-consistently. In particular, the dimensionless voltage is related to $\omega$ by $\langle \dot{\varphi} \rangle = \langle V \rangle / V_0 = \omega$. The ratio $g/\omega$ must be less than unity for running states, otherwise $\dot{\varphi}$ would reach zero. Similarly for the oscillator we employ the ansatz
\begin{equation}
    a = A \sin (\omega \tau + \theta),
\label{ansatz2}
\end{equation}
\noindent where $A$ and $\theta$ are also to be determined self-consistently. Substituting Eqs. (\ref{ansatz1}) and (\ref{ansatz2}) into Eqs. (\ref{rcsj}) and (\ref{mech}) we obtain a set of algebraic equations for the unknown parameters as functions of $\omega$. From these equations we obtain

\begin{eqnarray}
    && g = \left(\frac{2\omega(i_\text{DC} - \beta_1 \omega)}{\beta_1 + \mu^2 \beta_2 \omega^3 (2\omega/Q)/\kappa}\right)^{1/2}_{},
\label{g}\\
    && A = \mu \beta_2 g \omega / \sqrt{\kappa},
\label{A}\\
    && \cos (\varphi_0) = -\frac{2\omega}{g}(i_\text{DC}^{}-\beta_1\omega),
\label{cosp0}
\end{eqnarray}
\noindent with $\kappa = [1-(1+\mu^2)\omega^2]^2 + (2\omega/Q)^2$. The physical significance of the phase $\varphi_0$ is demonstrated by the fact that the time averaged supercurrent is given by $\langle \sin (\varphi) \rangle = -(g/\omega)\cos (\varphi_0)$. In fact Eq. (\ref{cosp0}) is obtained by simply performing the time average of the current conservation equation (\ref{rcsj}). These analytical solutions directly fulfil the energy balance equation (\ref{energyeq3}).

The DC voltage in the steady state is given by $\omega$ itself, and is found as a solution to the following eighth order polynomial equation
\begin{widetext}
\begin{equation}
\resizebox{.95\hsize}{!}{
$F(\omega) = 2 \omega (\beta_1 \omega - i_\text{DC}^{})\left[2 \mu^2 \beta_2 \omega^3 \left(\beta_1 \frac{2\omega}{Q} + \beta_2 \omega [1-\omega^2(1+\mu^2)]\right) + (\beta_1^2 + \beta_2^2 \omega^2)\kappa + (\mu^2 \beta_2 \omega^3)^2 \right] + \beta_1 \kappa + \mu^2 \beta_2 \omega^3 \frac{2\omega}{Q} =0\, .$
}
\label{poly}
\end{equation}
\end{widetext}
\noindent For $\omega<0$ and $\omega>i_\text{DC}^{}/\beta_1$, the polynomial $F(\omega)$ is strictly positive. As such, there must be an even number of real roots in the interval $(0,i_\text{DC}^{}/\beta_1)$. There turn out to be solutions for which the polynomial derivative is negative, $F'(\omega) < 0$. Such solutions give a voltage that decreases with increasing current and will be discarded as unphysical. All other real roots describe physical states that the system may enter with appropriate initial conditions.

\begin{figure}
	\includegraphics[width=75mm]{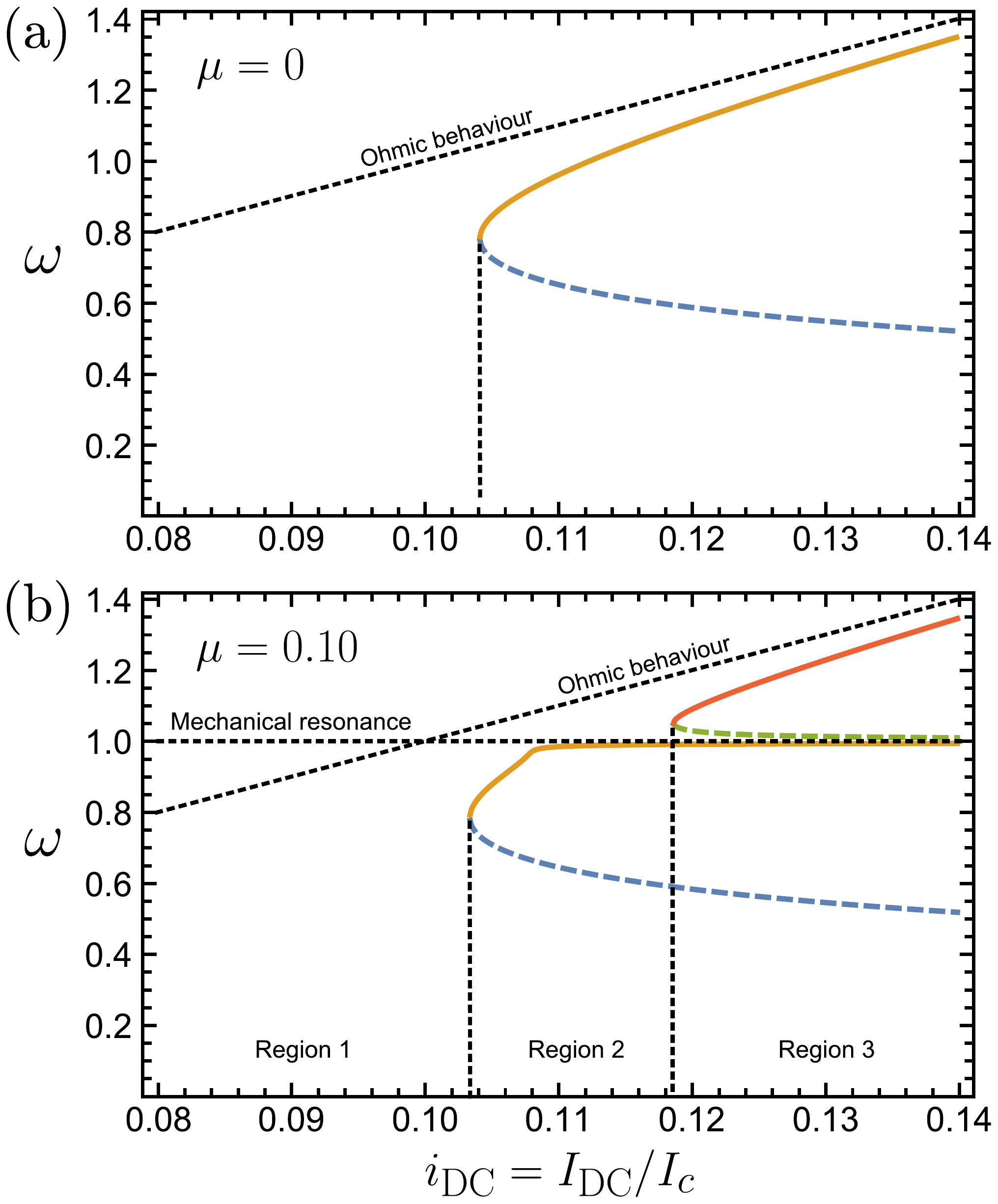}
	\caption{Real solutions of Eq. (\ref{poly}) as a function of the current $I_\text{DC}^{}$ for $\beta_1 = 0.1$, $\beta_2 = 2$, $Q = 10^3$, for (a) the uncoupled case $\mu=0$ and (b) the coupled case $\mu=0.10$. Unphysical solutions are shown as coloured dashed lines while the black dotted lines are guides for the eye to distinguish the various regimes of solutions. In the uncoupled case, the only physical solution approaches the Ohmic behaviour $\omega = i_\text{DC}^{}/\beta_1$. In the coupled case, the Ohmic solution still exists for high currents (red line in region 3), but for lower currents a new solution appears, which develops into a plateau at the mechanical resonance $\omega \approx 1$ (orange line in region 2). Above the critical coupling $\mu>\bar{\mu}$, the plateau solution becomes unstable and gives way to the trapped state with zero voltage. In both cases, below the retrapping current $I_\text{r}$ there are no physical solutions in the running state (region 1).}
	\label{analytical}
\end{figure}

Fig.\,\ref{analytical} shows the dependence of $\omega$ on $i_\text{DC}^{}$ leading to the I-V curve. In the uncoupled case $\mu = 0$ (panel (a)), there is only one physical solution (orange line) which approaches the Ohmic behaviour $\omega \approx i_\text{DC}^{}/\beta_1$ for large currents and vanishes below the retrapping current. The coupled case $\mu \neq 0$ (panel (b)), however, displays three distinct regions of solutions depending on the current $i_\text{DC}$. The low current regime of trapped states, i.e. region 1 in Fig.\,\ref{analytical}(b) where $i_\text{DC}$ is smaller than the retrapping current, is essentially unchanged from the uncoupled case. Above the retrapping current there exist two types of dynamics depending on whether the system is on or off the mechanical resonance. In region 2, where the system is close to resonance, mechanical oscillations are strong locking the overall dynamics to the resonant frequency $\omega_0$ (orange line). As a result, $\omega \approx 1$ regardless of $i_\text{DC}$ forming a Shapiro-like plateau, in agreement with the numerical analysis (cf. curves II and III in Fig.\,\ref{numerical}(b)). From the energy balance equation (\ref{energyeq3}), the energy dissipated mechanically in this region amounts to $i_\text{DC} - \beta_1$ which increases sharply upon entering the plateau. 

In region 3 there are two possible physical solutions: the Shapiro plateau solution (orange line) extending into this region, together with an Ohmic solution (red line) for which the electronic and mechanical subsystems are essentially uncoupled. In the latter, the energy is mainly dissipated electronically. Which state the system enters in this region can only be determined by initial conditions. In our scheme with the decreasing current path, the initial condition naturally realises the Ohmic solution. This explains the jump from the Ohmic solution in region 3 to the plateau solution in region 2 as $i_\text{DC}$ decreases for weak coupling with $\mu<\bar{\mu}$ (cf. Fig.\,\ref{numerical}).

As revealed in the numerical analysis, for strong coupling with $\mu>\bar{\mu}$ the system gets retrapped without forming a plateau (cf. curve IV in Fig.\,\ref{numerical}(b)). To understand this, we have performed a standard stability analysis, in which we perturb the system slightly off the plateau solution and study whether the perturbation grows or decays. We found that the plateau solution (orange line in Fig.\,\ref{analytical}(b)) is unstable if $\mu>\bar{\mu}$, leading to the Ohmic solution in region 3 as the only physically relevant one for strong electromechanical coupling.

\begin{figure}
	\includegraphics[width=85mm]{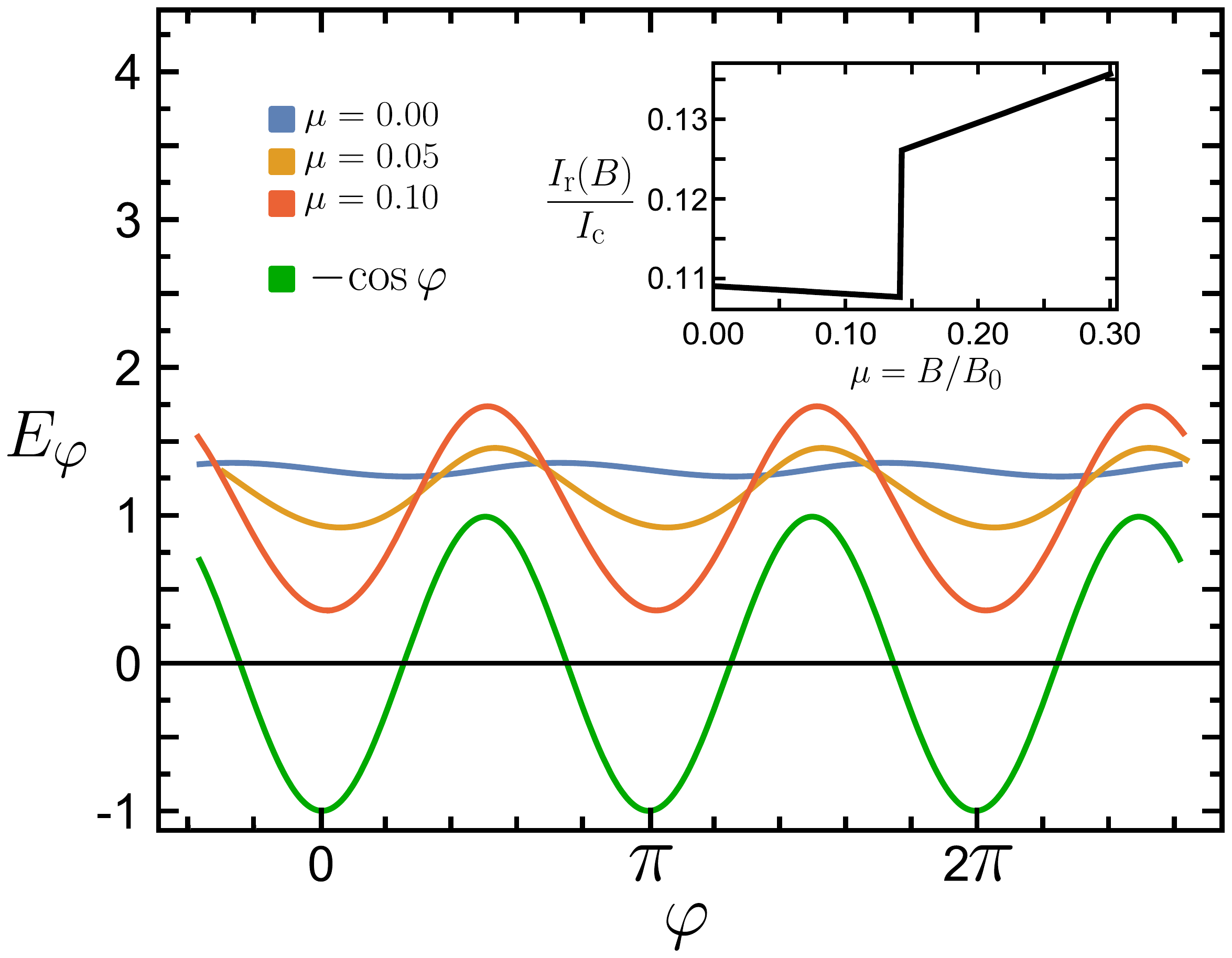}
	\caption{Evolution of the electronic energy $E_\varphi$ (in units $\hbar I_\text{c} / 2 e$) for numerical simulations with $\beta_1=0.1$, $\beta_2=2$, $Q=10^3$ and three different values of coupling $\mu$ ( $i_\text{DC}$ is chosen to be the highest current on the plateau solution where the mechanical oscillations are strongest). As coupling is increased, more energy is subtracted from the electronic system due to mechanical oscillations so that $E_\varphi$ decreases on average. When $\mu$ exceeds the critical value $\bar{\mu}$, $\langle E_\varphi \rangle < 1$ and the system retraps, as confirmed by the numerical solution. The inset shows the analytical prediction of the dependence of the retrapping current $I_\text{r}(B)$ on the coupling strength $\mu$.}
	\label{energy}
\end{figure}

A physical argument for understanding the retrapping arises from energy considerations. For the system to become trapped the electronic energy $E_{\varphi}^{}$ in Eq. (\ref{energyeq}) (in units $\hbar I_\text{c} / 2 e$) must be less than the maximum of the potential barrier $-\cos \varphi$, i.e. $\langle E_{\varphi}^{}\rangle\leq 1$. In the uncoupled case ($\mu=0$), mechanical oscillations are absent and the solution to Eq. (\ref{energyeq2}) is Ohmic ($\dot{\varphi}=i_\text{DC}^{}/\beta_1$). Here $E_{\varphi}^{}$ coincides with the total energy and is essentially constant. By demanding that the time derivative of the total energy in Eq. (\ref{energyeq2}) averages to zero over one period the critical condition for retrapping ($\langle E_{\varphi}^{}\rangle = 1$) yields an expression for the retrapping current $I_\text{r} \approx 4 I_\text{c}/\pi \sqrt{\beta_\text{c}}$. A similar argument can be employed in the coupled case ($\mu\neq 0$) except that the electronic energy $E_\varphi$ differs from the total energy $E$ due to energy being transferred to mechanical oscillations. As a consequence the system is retrapped at much higher currents than in the uncoupled case. Fig.\,\ref{energy} shows how $E_{\varphi}^{}$ decreases on average upon increasing the coupling strength. As $\mu$ approaches $\bar{\mu}$ the critical condition $\langle E_\varphi \rangle = 1$ is met. This simple condition along with the full analytical solution may be used to predict the dependence of the retrapping current $I_\text{r}(B)$ on the magnetic field, as shown in the inset of Fig.\,\ref{energy}. The figure shows that this condition suffices to reproduce the large increase in the retrapping current at a critical value of magnetic field $\bar{\mu}\approx 0.14$ which agrees well with the numerically found value $\bar{\mu} \approx 0.12$ (Fig.\,\ref{numerical}). The inset of Fig.\,\ref{energy} shows a decrease in $I_\text{r}(B)$ for $\mu<\bar{\mu}$ which is not observed in the numerical analysis. This is due to a limitation of the analytical model for very low currents where the fluctuations in voltage become comparable to the average value i.e. $g/\omega \sim 1$.

\subsection{Estimating the NEMS parameters from the I-V curve}

The I-V curve of the suspended Josephson weak link in the conventional DC current bias setup yields the resonant frequency $\omega_{0}^{}$ as well as the quality factor $Q$ of the resonator. In fact the resonant frequency $\omega_0^{}$ can be determined by simply measuring the voltage of the Shapiro plateau as the two are related by $V_0^{}=\hbar\omega_0/2e$. In experiments, to check that this feature is genuinely related to the mechanical resonance, one can modulate $\omega_0^{}$ by means of a DC back-gate voltage and detect the corresponding change in $V_0^{}$ from the I-V curve.

A useful estimate of $Q$ may be obtained by measuring the critical coupling $\bar{\mu}$ as may be demonstrated by considering the energy argument for retrapping. For a given $\mu$, mechanical oscillations are strongest at the highest current state on the plateau $i=i_{2/3}$, i.e. where the two regions 2 and 3 of Fig.\,\ref{analytical}(b) meet. The critical condition is therefore satisfied when $\mu = \bar{\mu}$ and $i=i_{2/3}$. For $\mu > \bar{\mu}$ the electronic subsystem has an energy $\langle E_\varphi \rangle < 1$ so that this retrapping occurs for all magnetic fields greater than the critical value. A crucial point is that the current $i_{2/3}$ itself depends on $\mu$ as can be seen in Fig.\,\ref{numerical}. The numerical results suggest the dominant scaling $(i_{2/3}-\beta_1) \propto \mu$ with a coefficient of order 1, as well as $\kappa \propto (2\omega / Q)^2$ and $\bar{\mu}^2 Q \beta_2 / 2 \beta_1 \gg 1$. With these assumptions the energy condition $\langle E_\varphi \rangle = 1$ yields $\bar{\mu}$ in terms of $Q$ as
\begin{equation}
    \bar{\mu} \approx \frac{\alpha}{Q^{1/3}},
\end{equation}
\noindent with $\alpha$ a constant of order unity. Our analysis for $Q=10^3$ yields $\alpha\simeq 1.4$. The critical coupling $\bar{\mu}$ thus scales as $Q^{-1/3}$ as noted previously. Since $\bar{\mu}$ may be measured directly from the I-V characteristic, this relation can be inverted to obtain an estimate for $Q$. With a high quality resonator with $Q = 10^6$ the predicted sudden retrapping could thus be observed with magnetic fields as small as $B \simeq 100 \, \si{mT}$.

\section{Discussion and conclusion}

In this paper we have demonstrated how the Josephson effect may be employed to activate and detect mechanical oscillations in an experimentally accessible DC current bias setup. By solving numerically and analytically the coupled equations describing the electronic and mechanical degrees of freedom of the system we have unveiled the non-trivial effects of the electromechanical coupling on the I-V characteristic. The appearance of a Shapiro-like plateau at weak coupling and the sudden retrapping of the system at large coupling allow for the exploration of the Josephson effect in a previously unstudied regime. Our analysis of the I-V curve reveals how to perform a purely DC measurement of both the resonant frequency and quality factor of the suspended resonator.

It has to be stressed that the experimental realisation of our proposal in the weak and strong electromechanical coupling regimes does not require any additional setup on top of the suspended weak link with superconducting contacts. Regarding the experimental feasibility of the measurement, in state of the art devices with resonance frequencies and quality factors of $\omega_0 \approx 10 \, \si{\giga \hertz}$ and $Q \approx 10^6$ \cite{peng,garcia}, the key features in the I-V characteristic appear at typical current and voltage scales of $\beta_1 I_\text{c} \approx 1 \, \si{\nano \ampere}$ and $V_0 \approx 3 \, \si{\micro \volt}$, with the crossover between weak and strong coupling occurring around a magnetic field of order $\bar{\mu} B_0 \approx 100 \, \si{\milli\tesla}$. Experiments show that this level of resolution can be obtained with current experimental setups \cite{herrero,cleuziou}. We have also investigated the effects of temperature by including a Johnson-Nyquist noise in our equations. To minimise the effects of temperature the thermal energy $k_B T$ must be less than the energy scale $\hbar I_\text{DC} / 2e$ associated with the current $I_\text{DC}$. Current state of the art CNT resonators with high resonant frequency $\omega_0 \approx 10 \, \si{\giga\hertz}$ \cite{peng,garcia} would allow the observation of the proposed effects for temperatures $T < 100 \, \si{\milli \kelvin}$, compatible with recent measurements on suspended Josephson junctions \cite{etakiSSO,kretinin}. 

\section*{Acknowledgments}

We thank Saverio Russo for stimulating discussions. Financial support from the Leverhulme Trust (Research Project Grant RPG-2015-101), and the Royal Society (International Exchange Grant Nr. IE140367, Newton Mobility Grants 2016/R1 UK-Brazil, and Theo Murphy Award TM160190) are gratefully acknowledged.

\section*{Appendix A: Numerical procedure and parameters}

The numerical solutions displayed in the paper were calculated using a fourth order Runge-Kutta method with the following parameters which are consistent with previously studied CNT devices \cite{peng,garcia,huttel,laird,moser,herrero,cleuziou}: $I_c = 10 \, \si{\nano \ampere}$, $R = 330 \, \si{\ohm}$, $\beta_c = 200$, $\omega_0 = 1 \, \si{\giga \hertz}$, $Q = 10^3$, $M = 10^{-20} \, \si{\kilogram}$, $L = 1 \, \si{\micro \metre}$. All other parameters may be derived from these, including those appearing in Eqs. (\ref{rcsj}) and (\ref{mech}), $\beta_1 = 0.1$, $\beta_2 = 2$ and the dimensional scales $V_0 = 0.3 \, \si{\micro \volt}$, $B_0 = 10 \, \si{\tesla}$, $x_0 = 10 \, \si{\pico \metre}$. While these are typical parameters, state of the art devices may have larger quality factors $Q = 10^6$ and resonance frequencies $\omega_0 = 10 \, \si{\giga \hertz}$ which allow measurements to be made at higher temperatures and lower magnetic fields, as discussed in the main text.

\section*{Appendix B: Finite temperature effects} 

Temperatures used in experiments must be below the critical temperature $T_c$ of the superconducting contacts. These could be realised with rigid nanostructures based on e.g. niobium nitride \cite{mizuno} or molybdenum rhenium \cite{molycontacts} with rather high $T_c \approx 10 \si{\kelvin}$. Even at these temperatures thermal currents and displacements may exceed the typical scales $I_\text{DC}^{}$ and $x_0$, disrupting the experimental signatures. We analysed temperature effects by including a Johnson-Nyquist noise current $\tilde{i}$ to our equations of motion with correlation function
\begin{equation}
\langle \tilde{i}(\tau) \tilde{i}(\tau') \rangle = 2 \beta_1 \frac{T}{T_J} \delta(\tau-\tau'),
\end{equation}
\noindent where $T_J$ is a temperature scale related to the Josephson energy $\hbar I_\text{c} / 2e = k_B T_J$. To minimise the effects of temperature the thermal energy $k_B T$ must be less than the energy scale $\hbar I_\text{DC} / 2e$ associated with the current $I_\text{DC}$ i.e. $T<\beta_1 T_J$. In terms of physical parameters, this condition reads
\begin{equation}
    k_B T< \left(\frac{\hbar}{2e}\right)^2 \frac{\omega_0}{R},
\end{equation}
\noindent so that experiments may be optimised using devices with low resistances and high mechanical resonance frequencies. For state of the art CNT devices with $\omega_0 \approx 10 \, \si{\giga \hertz}$ \cite{peng,garcia} and $R \approx 1 \, \si{\kilo \ohm}$ \cite{cleuziou,herrero} we find $T < 100 \, \si{\milli \kelvin}$.


\begin{thebibliography}{1}

    \bibitem{josephson} B. D. Josephson, 
    \emph{Possible New Effects in Superconductive Tunnelling}, \href{https://doi.org/10.1016/0031-9163(62)91369-0}
    {Phys. Lett. \textbf{1}, 251 (1962)}.

    \bibitem{shapiro} S. Shapiro,
    \emph{Josephson Currents in Superconducting Tunneling: The Effect of Microwaves and Other Observations},
    \href{https://doi.org/10.1103/PhysRevLett.11.80}
    {Phys. Rev. Lett. \textbf{11}, 80 (1963)}.
    
    \bibitem{coonfiske} D. D. Coon and M. D. Fiske,
    \emph{Josephson ac and Step Structure in the Supercurrent Tunneling Characteristic},
    \href{https://doi.org/10.1103/PhysRev.138.A744}
    {Phys. Rev. \textbf{138}, A744 (1965)}.
    
    \bibitem{baberschke} K. Baberschke, K. D. Bures, and S. E. Barnes,
    \emph{ESR in Situ with a Josephson Tunnel Junction},
    \href{http://dx.doi.org/10.1103/PhysRevLett.53.98}
    {Phys. Rev. Lett. \textbf{53}, 98 (1984)}.

    \bibitem{zhou} X. Zhou and A. Mizel,
    \emph{Nonlinear Coupling of Nanomechanical Resonators to Josephson Quantum Circuits},
    \href{https://doi.org/10.1103/PhysRevLett.97.267201}
    {Phys. Rev. Lett. \textbf{97}, 267201 (2006)}.

    \bibitem{buks} E. Buks and M. P. Blencowe,
    \emph{Decoherence and Recoherence in a Vibrating rf SQUID},
    \href{https://doi.org/10.1103/PhysRevB.74.174504}
    {Phys. Rev. B \textbf{74}, 174504 (2006)}.
    
    \bibitem{blencowe} M. P. Blencowe and E. Buks,
    \emph{Quantum Analysis of a Linear dc SQUID Mechanical Displacement Detector},
    \href{https://doi.org/10.1103/PhysRevB.76.014511}
    {Phys. Rev. B \textbf{76}, 014511 (2007)}.
    
    \bibitem{buks2} E. Buks, E. Segev, S. Zaitsev, B. Abdo, and M. P. Blencowe,
    \emph{Quantum Nondemolition Measurement of Discrete Fock States of a Nanomechanical Resonator},
    \href{http://doi.org/10.1209/0295-5075/81/10001}
    {Europhys. Lett. \textbf{81}, 10001 (2008)}.
    
    \bibitem{zhu} J.-X. Zhu, Z. Nussinov, and A. V. Balatsky,
    \emph{Vibration-Mode-Induced Shapiro Steps and Back Action in Josephson Junctions},
    \href{https://doi.org/10.1103/PhysRevB.73.064513}
    {Phys. Rev. B \textbf{73}, 064513 (2006)}.
    
    \bibitem{sonne1} G. Sonne, R. I. Shekhter, L. Y. Gorelik, S. I. Kulinich, and M. Jonson,
    \emph{Superconducting Pumping of Nanomechanical Vibrations},
    \href{https://doi.org/10.1103/PhysRevB.78.144501}
    {Phys. Rev. B \textbf{78}, 144501 (2008)}. 

    \bibitem{sonne2} G. Sonne, M. E. Pe{\~n}a-Aza, L. Y. Gorelik, R. I. Shekhter, and M. Jonson,
    \emph{Cooling of a Suspended Nanowire by an ac Josephson Current Flow},
    \href{https://doi.org/10.1103/PhysRevLett.104.226802}
    {Phys. Rev. Lett. \textbf{104}, 226802 (2010)}.
    
    \bibitem{sonne3} G. Sonne and L. Y. Gorelik,
    \emph{Ground-State Cooling of a Suspended Nanowire Through Inelastic Macroscopic Quantum Tunneling in a Current-Biased Josephson Junction},
    \href{https://doi.org/10.1103/PhysRevLett.106.167205}
    {Phys. Rev. Lett. \textbf{106}, 167205 (2011)}.
    
    \bibitem{padurariu} C. Padurariu, C. J. H. Keijzers, and Y. V. Nazarov,
    \emph{Effect of Mechanical Resonance on Josephson Dynamics},
    \href{https://doi.org/10.1103/PhysRevB.86.155448}
    {Phys. Rev. B \textbf{86}, 155448 (2012)}.
    
    \bibitem{marchenkov} A. Marchenkov, Z. Dai, B. Donehoo, R. N. Barnett, and U. Landman,
    \emph{Alternating Current Josephson Effect and Resonant Superconducting Transport Through Vibrating Nb Nanowires},
    \href{https://doi.org/10.1038/nnano.2007.218}
    {Nat. Nanotechnol. \textbf{2}, 481 (2007)}.
    
    \bibitem{keijzers} H. Keijzers,
    \emph{Josephson Effects in Carbon Nanotube Mechanical Resonators and Graphene},
    Ph.D. thesis, Delft University of Technology (2012).

    \bibitem{etaki} B. H. Schneider, S. Etaki, H. S. J. van der Zant, and G. A. Steele,
    \emph{Coupling Carbon Nanotube Mechanics to a Superconducting Circuit},
    \href{https://doi.org/10.1038/srep00599}
    {Sci. Rep. \textbf{2}, 599 (2012)}.
    
    \bibitem{etakiSSO} S. Etaki, F. Konschelle, Y. M. Blanter, H. Yamaguchi, and H. S. J. van der Zant,
    \emph{Self-Sustained Oscillations of a Torsional SQUID Resonator Induced by Lorentz-Force Back-Action},
    \href{https://doi.org/10.1038/ncomms2827}
    {Nat. Commun. \textbf{4}, 1803 (2013)}.
    
    \bibitem{kretinin} A. Kretinin, A. Das, and H. Shtrikman,
    \emph{The Self-Actuating InAs Nanowire-Based Nanoelectromechanical Josephson Junction},
    \href{https://arxiv.org/abs/1303.1410v2}
    {arXiv:1303.1410v2}.
    
    \bibitem{koch} J. Koch and F. von Oppen,
    \emph{Franck-Condon Blockade and Giant Fano Factors in Transport through Single Molecules},
    \href{https://doi.org/10.1103/PhysRevLett.94.206804}
    {Phys. Rev. Lett. \textbf{94}, 206804 (2005)}.
    
    \bibitem{koch2} J. Koch, F. von Oppen, and A. V. Andreev,
    \emph{Theory of the Franck-Condon Blockade Regime},
    \href{https://doi.org/10.1103/PhysRevB.74.205438}
    {Phys. Rev. B \textbf{74}, 205438 (2006)}.
    
    \bibitem{sapmaz} S. Sapmaz, P. Jarillo-Herrero, Y. M. Blanter, C. Dekker, and H. S. J. van der Zant,
    \emph{Tunneling in Suspended Carbon Nanotubes Assisted by Longitudinal Phonons},
    \href{https://doi.org/10.1103/PhysRevLett.96.026801}
    {Phys. Rev. Lett. \textbf{96}, 026801 (2006)}.
    
    \bibitem{leturcq} R. Leturcq, C. Stampfer, K. Inderbitzin, L. Durrer, C. Hierold, E. Mariani, M. G. Schultz, F. von Oppen, and K. Ensslin,
    \emph{Franck-Condon Blockade in Suspended Carbon Nanotube Quantum Dots},
    \href{https://doi.org/10.1038/nphys1234}
    {Nature Phys. \textbf{5}, 327 (2009)}.
    
    \bibitem{likharev} K. K. Likharev,
    \emph{Superconducting Weak Links},
    \href{https://doi.org/10.1103/RevModPhys.51.101}
    {Rev. Mod. Phys. \textbf{51}, 101 (1979)}.
    
    \bibitem{peng} H. B. Peng, C. W. Chang, S. Aloni, T. D. Yuzvinsky, and A. Zettl,
    \emph{Ultrahigh Frequency Nanotube Resonators},
    \href{https://doi.org/10.1103/PhysRevLett.97.087203}
    {Phys. Rev. Lett. \textbf{97}, 087203 (2006)}.
    
    \bibitem{garcia} D. Garcia-Sanchez, A. San Paulo, M. J. Esplandiu, F. Perez-Murano, L. Forr{\'o}, A. Aguasca, and A. Bachtold,
    \emph{Mechanical Detection of Carbon Nanotube Resonator Vibrations},
    \href{https://doi.org/10.1103/PhysRevLett.99.085501}
    {Phys. Rev. Lett. \textbf{99}, 085501 (2007)}.
    
    \bibitem{huttel} A. K. H{\"u}ttel, G. A. Steele, B. Witkamp, M. Poot, L. P. Kouwenhoven, and H. S. J. van der Zant,
    \emph{Carbon Nanotubes as Ultrahigh Quality Factor Mechanical Resonators},
    \href{https://doi.org/10.1021/nl900612h}
    {Nano Lett. \textbf{9}, 2547 (2009)}.
    
    \bibitem{laird} E. A. Laird, F. Pei, W. Tang, G. A. Steele, and L. P. Kouwenhoven,
    \emph{A High Quality Factor Carbon Nanotube Mechanical Resonator at 39 GHz},
    \href{https://doi.org/10.1021/nl203279v}
    {Nano Lett. \textbf{12}, 193 (2012)}.
    
    \bibitem{moser} J. Moser, A. Eichler, J. G{\"u}ttinger, M. I. Dykman, and A. Bachtold,
    \emph{Nanotube Mechanical Resonators with Quality Factors of up to 5 Million},
    \href{https://doi.org/10.1038/nnano.2014.234}
    {Nat. Nanotechnol. \textbf{9}, 1007 (2014)}.
    
    \bibitem{bunch} J. S. Bunch, A. M. van der Zande, S. S. Verbridge, I. W. Frank, D. M. Tanenbaum, J. M. Parpia, H. G. Craighead, and P. L. McEuen,
    \emph{Electromechanical Resonators from Graphene Sheets},
    \href{https://doi.org/10.1126/science.1136836}
    {Science \textbf{315}, 490 (2007)}.
    
    \bibitem{chen} C. Chen, S. Rosenblatt, K. I. Bolotin, W. Kalb, P. Kim, I. Kymissis, H. L. Stormer, T. F. Heinz, and J. Hone,
    \emph{Performance of Monolayer Graphene Nanomechanical Resonators with Electrical Readout},
    \href{https://doi.org/10.1038/nnano.2009.267}
    {Nat. Nanotechnol. \textbf{4}, 861 (2009)}.
    
    \bibitem{tmd} N. Morell, A. Reserbat-Plantey, I. Tsioutsios, K. G. Sch{\"a}dler, F. Dubin, F. H. L. Koppens, and A. Bachtold,
    \emph{High Quality Factor Mechanical Resonators Based on WSe2 Monolayers},
    \href{https://doi.org/10.1021/acs.nanolett.6b02038}
    {Nano Lett. \textbf{16}, 5102 (2016)}.
    
    \bibitem{tinkham} M. Tinkham,
    \emph{Introduction to Superconductivity}
    (Dover Publications, New York, 2004) 
    
    \bibitem{heersche} H. B. Heersche, P. Jarillo-Herrero, J. B. Oostinga, L. M. K. Vandersypen, and A. F. Morpurgo,
    \emph{Bipolar Supercurrent in Graphene},
    \href{https://doi.org/10.1038/nature05555}
    {Nature \textbf{446}, 56 (2007)}.
    
    \bibitem{mizuno} N. Mizuno, B. Nielsen, and X. Du,
    \emph{Ballistic-Like Supercurrent in Suspended Graphene Josephson Weak Links},
    \href{https://doi.org/10.1038/ncomms3716}
    {Nat. Commun. \textbf{4}, 2716 (2013)}.
    
    \bibitem{herrero} P. Jarillo-Herrero, J. A. van Dam, and L. P. Kouwenhoven,
    \emph{Quantum Supercurrent Transistors in Carbon Nanotubes},
    \href{https://doi.org/10.1038/nature04550}
    {Nature \textbf{439}, 953 (2006)}.
    
    \bibitem{cleuziou} J.-P. Cleuziou, W. Wernsdorfer, V. Bouchiat, T. Ondar{\c c}uhu, and M. Monthioux,
    \emph{Carbon Nanotube Superconducting Quantum Interference Device},
    \href{https://doi.org/10.1038/nnano.2006.54}
    {Nat. Nanotechnol. \textbf{1}, 53 (2006)}.
    
    \bibitem{molycontacts} M. Aziz, D. C. Hudson, and S. Russo,
    \emph{Molybdenum-Rhenium Superconducting Suspended Nanostructures},
    \href{https://doi.org/10.1063/1.4883115}
    {Appl. Phys. Lett. \textbf{104}, 233102 (2014)}.

\end{thebibliography}
\end{document}